\documentclass{article}
\usepackage[preprint]{neurips_2022}

\usepackage[utf8]{inputenc} 
\usepackage[T1]{fontenc}    
\usepackage[hidelinks]{hyperref}       
\usepackage{url}            
\usepackage{booktabs}       
\usepackage{amsfonts}       
\usepackage{nicefrac}       
\usepackage{microtype}      
\usepackage{xcolor}         
\usepackage{multirow}
\usepackage{graphicx}
\usepackage{amsmath}
\usepackage{float}

\title{Interaction Order Prediction for Temporal Graphs}

\author{%
  Nayana Bannur \\
  Machine Learning Department \\
  Carnegie Mellon University \\
  \texttt{nbannur@cs.cmu.edu} \\
  \And
  Mashrin Srivastava \\
  Machine Learning Department \\
  Carnegie Mellon University \\
  \texttt{mashrins@cs.cmu.edu} \\
  \And
  Harsha Vardhan \\~
  Machine Learning Department \\
  Carnegie Mellon University \\
  \texttt{vvl@cs.cmu.edu} \\
}

\begin{document}

\maketitle

\section{Introduction}
\label{sec:intro}
Link prediction in graphs is a task that has been widely investigated. It has been applied in various domains such as knowledge graph completion \citep{lp-for-kg}, content/item recommendation \citep{jodie}, social network recommendations \citep{lp-for-social-network} and so on. The initial focus of most research was on link prediction in static graphs. However, there has recently been abundant work on modeling temporal graphs, and consequently one of the tasks that has been researched is link prediction in temporal graphs. 

Different types of temporal link prediction tasks have been proposed. One of the most common formulations is binary classification where given a snapshot of a graph at time $t$ the goal is to predict the links formed at time $t+1$. However, most of the existing work does not focus on the order of link formation, and only predicts the existence of links. Predicting the sequential order of node interactions could be useful for recommendation systems \citep{bert4rec}, making predictions for social network and other domains, for both early predictions and long-term predictions. In this project, we aim to predict the order of node interactions. As a starting point, we target a specific task which is interaction order prediction (IOP) \citep{tat}. 

\paragraph{Problem Formulation} 
\label{para:intro-prob-form}
Define an undirected temporal graph $G = (V, E, T)$ with nodes $V = \{v_1, \dots, v_N\}$, edges $E$ and timestamps $T = \{T_{ij}\}$, such that $T_{ij}$ is a list of timestamps at which nodes $i$ and $j$ interact. Given $G$ and a set of $n$ nodes $\{v_1, \dots, v_n\} \in V$, the task is to predict the the interaction order of the nodes. Interaction order for an $n$-node set is the order in which edges are formed, only considering the first interaction for any pair of nodes. For example, for a clique $\{v_1, v_2, v_3\}$ there are six interaction orders possible such as $12 \to 23 \to 13$, $13 \to 12 \to 23$, etc.

\paragraph{Data} 
\label{para:intro-data}
The task can be performed using any graph dataset consisting of a set of observations representing node interactions with timestamps. In other words, the dataset consists of a list of tuples of the form $(v_i, v_j, t)$ where $v_i, v_j \in V$ and $t$ indicates the timestamp at which they interact. The datasets used are described in detail in Section \ref{subsec:data}.

\section{Background}
\label{sec:background}
The interaction order prediction problem was proposed in \citep{tat}, where the order of interaction of a set of three nodes was predicted as a multi-class classification problem over the permutations of nodes. We evaluated existing work for this task. A set of existing GNN models were used as baselines including GCN \citep{gcn}, GAT \citep{gat}, GraphSAGE \citep{graphsage}, TAGCN \citep{tagcn}, DE-GNN \citep{degnn} and TAT \citep{tat}. The baselines included models which were developed for static graphs (for example, GCN), as well as models which were developed for temporal graphs (for example, TAT).
All of these models can be used for learning node representations. Each model was combined with a classification network and trained end-to-end on node sets to predict their interaction order. All baselines were trained for node sets of size three and four, using the SMS-A\citep{smsa} and CollegeMsg\citep{collegemsg} datasets. The results were evaluated using a set of metrics consisting of accuracy, AUC, BLEU-3\citep{bleu} and Kendall's rank correlation coefficient\citep{kendall}. 

The TAT model performed the best out of the baselines. However, the metrics for the baselines were not very high. Also, it was observed that for four-node sets the performance dropped drastically for all baselines. This is because the number of target classes was very large for the amount of data available. Our takeaway was that the multi-class classification formulation was not scalable. We were motivated to find a different approach to the IOP task which could both improve the existing results and also make the solution scalable to larger node sets. 

\section{Related Work}
\label{sec:rel-work}

\subsection{Learning for temporal graphs}
Many methods have emerged for learning representations for temporal graphs. 
Temporal Graph Networks (TGN) \citep{tgn} maintain memory for each node and update this memory using messages propagated from neighbors. JODIE \citep{jodie} focuses on predicting node embeddings trajectories in temporal graphs. The Temporal ATtention network (TAT) \citep{tat} captures temporal information via a time encoder which discretizes interaction timestamps and encodes them similar to position encodings in transformers. The temporal information is then used to compute attention weights for neighborhood aggregation. Temporal Graph Attention (TGAT) \citep{tgat} is a similar approach where a time-varying representation is computed for each node and self-attention weights are computed to perform neighborhood aggregation.  Neighborhood Extended Dynamic Graph Neural Network (NEDGNN) \citep{EDGNN} uses a temporal attention propagation module which uses self-attention mechanism on $n$-hop neighbors for information propagation and a FIFO message box for time efficiency. Another one is the Instant Graph Neural Network (InstantGNN) \citep{InstantGNN} which is an incremental computation approach for the graph representation matrix of dynamic graphs. It uses instant updates on the representations and instant predictions along with an adaptive training strategy which boosts the performance. We use some of these approaches as baselines.

\subsection{Temporal link prediction}
There are various ways of defining temporal link prediction. The primary method is binary classification where given a snapshot of the graph, the task is to predict future links. A model is trained on historical information until time $t$ and then predicts whether a given pair of nodes will form a link in the future or predicts the adjacency graph at $t+1$ \citep{tlp_bc1}. 
Another method is to predict the adjacency matrix of a graph at time $t+1$, given a series of snapshots until time $t$ of a temporal graph \citep{tpl_bc2}.
Recent methods propose using node ranking for link prediction of time-evolving network \citep{Wu2020LinkPO}. DLP-LES \citep{Selvarajah2020DynamicNL} is a novel framework which learns the transitional patterns of a given dynamic network using common neighbors based on subgraphs of a target link. It uses heuristic features of the subgraph extracted using CNN-LSTM to obtain additional information. E-LSTM-D \citep{Chen2019ELSTMDAD} is a unified framework which uses a LSTM, together with an encoder–decoder architecture for link prediction in dynamic networks. It automatically learns structural and temporal features for link prediction in the future. All these methods either predict the existence of links without ordering, or only focus on predicting the single next link. Another formulation is predicting the exact timestamp of the next interaction between two nodes \citep{tmp_timestamp}. The interaction order prediction problem is proposed in \citep{tat}, where the order of interaction of a set of three nodes is predicted as a multi-class classification problem over the permutations of nodes. Although this work predicts an ordered sequence of interactions, the work is limited to node triplets. This approach leads to increase in complexity as the size of the node set increases since the number of target classes grows factorially with the target sequence size. The interaction order problem could be viewed as sequence prediction rather than classification, in order to scale to larger node sets.

\subsection{Sequence prediction}
Sequence prediction is a common task across many domains. In natural language processing, sequence prediction is found in tasks such as machine translation \citep{translation}, captioning \citep{captioning}, summarization \citep{summarization} and so on. Another domain which bears some similarities is time series forecasting. However, most forecasting techniques are not suitable for predicting a series of integers from a fixed set. There is some relevant work in other domains as well. For example in recommendation systems, BERT representations \citep{BERT} using bidirectional self-attention have been used to model the interactions for sequential recommendation\citep{bert4rec}. Another example is clinical event sequence prediction in which neural models auto-regressive learn personalized patient-specific representations to predict future clinical events \citep{clinicalevent}. 
We see that sequence prediction is typically performed in an auto-regressive manner which motivates us to adopt a similar approach. 
There has been some work on sequence prediction using graph data as well. Generative Link Sequence Modeling (GLSM) \citep{glsm} uses temporal link patterns in a sequence modeling framework to generate a probability distribution over the possible future links. It further proposes self tokenization for generalization beyond raw link sequences as raw links are transformed as abstract aggregation tokens.

\subsection{Evaluation}
Temporal link prediction is typically evaluated using standard metrics such as accuracy, AUC etc. when it is posed as a binary classification problem. When it is formulated as sequence prediction, evaluation techniques of similar tasks in other domains can be investigated and adapted. Language generation is a sequence prediction problem where metrics such as BLEU \citep{bleu} and METEOR \citep{meteor} are used. While there are a large number of evaluation metrics in NLP, only model-free metrics are suitable to use outside the domain of text. Further, since predicting interaction order does not require constraints such as brevity penalties and others, the number of metrics that are relevant is further limited. Ranking models generate an ordered ranking and metrics such as the Kendall \citep{kendall} and Spearman's \citep{spearman} rank correlation coefficients are used to measure rank correlation. Although multi-step forecasting is a similar problem in some ways, its metrics such as RMSE, MAPE, etc. are not suitable.

\section{Methods}
\label{sec:methods}

In the following sections, we describe the methods implemented beyond the baselines. In an attempt to reformulate the link prediction task we propose the sequence prediction based approach, building off of the TAT model \citep{tat}. We also design an additional model which focuses on making predictions at specific time horizons. While these two approached are based on the TAT model, we also design a model based off of JODIE \citep{jodie} which learns embedding trajectories, in order to have more control over the temporal changes in embeddings. 

\subsection{Sequence Prediction}
The baseline consisted of the TAT encoder with a classification network. We converted the multi-class classification problem into a sequence prediction problem by replacing the classification network with an RNN decoder. Our primary motivation was to take successful approaches from other domains such as NLP and introduce them into temporal graph modeling.

The TAT encoder generates embeddings for the node-set in consideration. These embeddings can be used to compute a context vector for the RNN, which is used to generate an output sequence corresponding to interaction order. The context vector can be created as the concatenation the embeddings, mean of the embeddings, etc.

The IOP problem pertains to predicting the correct permutation of interactions. While a RNN-based decoder can generate a sequence of interactions, it does not necessarily generate a permutation. The RNN model may generate repeated tokens and may exclude certain tokens in the output vocabulary. However, we want our sequence to contain all tokens in the output vocabulary exactly once each. In order to impose these constraints, we performed post-processing on the logits generated by the RNN. Techniques such as repetition penalties \citep{repetition-penalty} are used to reduce repetition. Since we wanted to have no repetitions we uses a technique which prevents repetition of n-grams, with $n=1$. This ensures that the output sequence is a permutation of the target sequence.

\subsection{Prediction at Timestep}
\label{subsec:pred-t}
The baseline model used a set of $n$ nodes and a sub-graph extracted around this node set as the input. It generated embeddings using these nodes and created a concatenated embedding as the classifier input. The classifier output was a label from ${n \choose 2}!$ classes. We modified the classifier to accept a timestep $t \in [1, \dots, {n \choose 2}]$, in addition to the concatenated node embeddings. This timestep corresponded to which interaction to predict, out of the sequence of interactions of length ${n \choose 2}$. We also modified the classifier to predict the two nodes which interacted out of the $n$ nodes, rather than a label corresponding to a specific permutation. The motivation behind this model was twofold. First, we wanted to experiment with changing the input and output representations to see if this would ease learning. Second, this model could be used analyse the usefulness of the embeddings in predicting different lengths of sequences and making predictions at a specific horizon (i.e. predicting the $t^{th}$ interaction).

\subsection{Dynamic Embeddings}
JODIE \citep{jodie} is a framework which predicts node embedding trajectories for bipartite temporal interaction networks. The model consists of RNN layers to learn node embeddings. It is able to update the embeddings with each interaction observed. It also learns embedding layers to project the node embeddings temporally. Due to its ability to manipulate embeddings which was not captured in the TAT encoder, we experimented with JODIE to generate embeddings.

JODIE is designed for heterogeneous graphs which capture user-item interactions. Since our work focuses on homogeneous graphs, we modified the model for homogeneous networks. We modified the framework to maintain a single embedding space for all users, rather than item embeddings and user embeddings. Further, for each interaction between users, we updated the embeddings for both users in the same way, since we consider the graph to be undirected.

The framework learns embeddings using the binary link prediction task. The embeddings were used as input to an MLP classifier to predict interaction order. Although the setup could have been modified for sequence prediction, we used an MLP classifier as a starting point. 

\section{Experiments}
\label{sec:experiments}

In this section we first discuss the datasets, experimental setup and metrics used for evaluation. We then describe the experiments related to each of the methods proposed in the previous section.

\subsection{Data}
\label{subsec:data}
The datasets we chose are: 
\begin{enumerate}
    \item SMS-A \citep{smsa}: This dataset consists of messages with timestamps sent between users, collected from a mobile phone operator.
    \item CollegeMsg \citep{collegemsg}: This dataset consists of messages with timestamps sent between users of an online social network. The dataset was collected from an online social community at the University of California, Irvine. 
\end{enumerate}

The dataset statistics are summarized in Table \ref{tab:data}.

\begin{table}[htb]
    \centering
    \begin{tabular}{cccccc}
    \toprule
    \multirow{2}{*}{Dataset} & \multirow{2}{*}{Vertices ($|V|$)} & \multirow{2}{*}{Edges ($|E|$)} & \multirow{2}{*}{Timestamps ($|T|$)} & \multicolumn{2}{c}{Complete sub-graphs} \\
    {} & {} & {} & {} & $n=3$ & $n=4$ \\
    \midrule
    SMS-A & 44430 & 53866 & 548182 & 3769 & 281 \\
    CollegeMsg & 1899 & 13838 & 59835 & 7135 & 1304 \\
    \bottomrule
    \end{tabular}
    \caption{Summaries of SMS-A and CollegeMsg datasets.}
    \label{tab:data}
\end{table}

For models using the TAT encoder, the models were trained on sub-graphs corresponding to node sets. For the IOP task, we used node sets which were complete sub-graphs of $n$ nodes. Pre-processing included finding max cliques of size $n$ and extracting sub-graphs corresponding to the k-hop neighborhood for each clique ($k=1$ was used). Table \ref{tab:data} reports the number of sub-graphs extracted. For models using the modified JODIE encoder, the entire graph was using for training rather than sub-graphs.

\subsection{Experimental Setup}
From the original datasets we constructed three specific datasets for our experiments - SMS-A with $n=3$, CollegeMsg with $n=3$ and CollegeMsg with $n=4$\footnote{We excluded SMS-A with $n=4$ since it contained too few samples}. We evaluated all the experiments using one or more of these datasets. 

The training-validation-test split used was 80\%-10\%-10\%. Since the type of data used for the TAT and JODIE based models differed, the data in the different splits were not exactly same, so the results from these experiments were not directly comparable. JODIE was trained using all interactions up to a time point.

The hyperparameters for the TAT model used were fixed (unless mentioned otherwise). The embedding size used was 128 and the context vector was created by concatenation. For JODIE, the hyperparameters used were set to the default values used in the original framework. 

Across all experiments, all models were trained for 50 epochs each. The test results corresponding to the best validation metrics were reported. 

\subsection{Metrics}
\label{subsec:metrics}

The metrics used for evaluation consist of standard classification metrics, model-free NLP metrics for sequence evaluation and non-parametric rank correlation coefficients. The following metrics were selected:

\begin{enumerate}
    \item Accuracy: Since the baselines were developed only for 3-node sets in the original work, one of the primary metrics reported was accuracy. In order to see how many exact predictions we are able to obtain, we continued to report accuracy. However, since accuracy is not an suitable metric for larger node sets, we also added other metrics.
    \item BLEU \citep{bleu}: 
    BLEU is one of the most common model-free evaluation metrics used to evaluate sequences in NLP. We chose BLEU-3 with no brevity penalty since the length of the predicted sequence and target are always same. Since the minimum length of the prediction sequence in our work was 3, BLEU-3 was more appropriate than BLEU-4 which is the most commonly used version. When predicting permutations, 1-grams do not have any contribution to the metric. However in some of our experiments the permutation restriction was not imposed hence we retained 1-gram precision in the metric. Note that BLEU-3 is equivalent to accuracy when predicting permutations of 3-node sequences (see Appendix \ref{app:bleu-acc} for more details).
    \item METEOR \citep{meteor}: METEOR is also a model-free NLP evaluation metric which is based on the harmonic mean of unigram precision and recall, with a higher weight assigned to recall.
    \item Kendall rank correlation coefficient \citep{kendall}:
    Kendall's rank correlation coefficient is a metric used to measure ordinal association of two sequences defined as:
    \[
    \tau = \frac{ \text{number of concordant pairs} - \text{number of discordant pairs}}{\text{number of pairs}}
    \]
    We chose this metric because it captures how many pairs of links were relatively ordered correctly and incorrectly.
    \item Spearman's rank correlation coefficient: The Spearman correlation coefficient is the Pearson correlation coefficient between rank variables. For variables $x$ and $y$ yt is given by:
    \[
    r = \frac{ \text{cov}(\text{R}(x), \text{R}(x))}{\sigma_{\text{R}(x)}\sigma_{\text{R}(y)}}
    \]
    where cov is the covariance, R represents the variable as a rank and $\sigma$ is standard deviation.
\end{enumerate}

Note that for both the correlation coefficients, we do not consider p-values because the p-values computed are only reasonable for large sequences (> 500) whereas the sequences considered in our work are of small length (3 or 4).

\subsection{Sequence Prediction Experiments}
We used a simple decoder RNN having a single GRU layer. In the main experiments, the context vector for the decoder was created by concatenating the embeddings of the node set. We evaluated the performance of the RNN without post-processing the logits (\texttt{TAT-sequence}) and with post-processing to enforce permutations (\texttt{TAT-sequence-perm}). We also varied hyperparameters such as the method of creating the context vector and hidden size.

\subsection{Prediction at Timestep Experiments}
We performed two types of experiments using the model described in Section \ref{subsec:pred-t}:
\begin{enumerate}
    \item \texttt{TAT-time-all}: We trained the model at all timesteps of the output sequence simultaneously. For example, for a node set $\{1,2,3\}$ with interaction order $[12, 13, 23]$, three input samples were created. If the concatenated node embedding is represented by $x$, the (input, output) of the samples were ([$x$, 1], [1, 1, 0]), ([$x$, 2], [1, 0, 1]) and ([$x$, 3], [0, 1, 1]). The motivation behind this model was to see whether the model performed better with different input and output representations. 
    \item \texttt{TAT-time-t}: We trained a separate model for individual timesteps. For example, for 3-node sets we trained three models \texttt{TAT-time-1}, \texttt{TAT-time-2} and \texttt{TAT-time-3}. This model was primarily used for diagnostic purposes, i.e. to understand the change in performance of the model as the time horizon of the prediction increases.
\end{enumerate}

\subsection{Dynamic Embeddings Experiments}
\label{subsec:exp-dyn-emb}
We conducted multiple experiments using our modified version of JODIE. The embeddings were trained using the binary link prediction task used in the original framework. The embeddings of the node set were concatenated and used as input for the classifier. The classifier was an MLP with a single hidden layer of size 128. Since this experiment was computationally expensive, we only ran it for the SMS-A dataset. The main experiments performed were:
\begin{enumerate}
    \item \texttt{dyn-emb}: The embeddings obtained at the end of training were directly used to train a classifier.
    \item \texttt{dyn-emb-projected}: The embeddings obtained at the end of training were projected forward in time before classification. This is because the interactions in the test set are at a later time and predicting them requires updated embeddings. During embedding training, the user embeddings were projected forward in time using the elapsed time from the previous interaction. The elapsed times of the training samples were scaled to have mean zero and unit variance. However during test time, since we would not know the elapsed times, we sampled values from a standard normal Gaussian as elapsed times, i.e. from the same distribution as the elapsed times in the training data. We sorted the samples such that the earlier interactions in the test set had a smaller elapsed time and later interactions had a larger elapsed time.
\end{enumerate}

\section{Results}

The results for the experiments are presented in this section. The original TAT model \citep{tat} was used as a baseline. The remaining baselines from our previous report are included in Appendix \ref{app:baselines} for reference. The metrics discussed in Section \ref{subsec:metrics} were used for evaluation.

\subsection{Sequence Prediction}

\begin{table}[htb]
    \centering
    \begin{tabular}{lrlccccc}
        \toprule 
        Dataset & $n$ & Model & Acc & BLEU-3 & METEOR & Kendall's $\tau$ & Spearman $\rho$\\
        \midrule
        \multirow[c]{3}{*}{SMS-A} & \multirow[c]{3}{*}{3} & \texttt{TAT-baseline} 
        & 0.237 & 0.237 & 0.725 & 0.167 & 0.192\\
        & & \texttt{TAT-sequence} & 0.205 & 0.205 & 0.677 & 0.050 & 0.057 \\
        & & \texttt{TAT-sequence-perm} & 0.219 & 0.219 & 0.707 & 0.060 & 0.069 \\
        \midrule
        \multirow[c]{3}{*}{CollegeMsg} & \multirow[c]{3}{*}{3} & \texttt{TAT-baseline} 
        & 0.415 & 0.415 & 0.746 & 0.311 & 0.331\\
        & & \texttt{TAT-sequence} & 0.300 & 0.300 & 0.684 & 0.171 & 0.217 \\
        & & \texttt{TAT-sequence-perm} & 0.363 & 0.363 & 0.725 & 0.252 & 0.283 \\
        \midrule
        \multirow[c]{3}{*}{CollegeMsg} & \multirow[c]{3}{*}{4} & \texttt{TAT-baseline} 
        & 0.000 & 0.134 & 0.688 & 0.067 & 0.087 \\
        & & \texttt{TAT-sequence} & 0.000 & 0.027 & 0.337 & 0.061 & 0.074\\
        & & \texttt{TAT-sequence-perm} & 0.000 & 0.083 & 0.676 & 0.063 & 0.071\\
        \bottomrule
    \end{tabular}
    \caption{Results for sequence prediction experiments using \texttt{TAT-sequence} and \texttt{TAT-sequence-perm}.}
    \label{tab:results-seq-pred}
\end{table}

Table \ref{tab:results-seq-pred} shows the result for the sequence prediction experiments. \texttt{TAT-sequence} had the lowest performance across metrics. This is expected since this model does not necessarily generated permutations, whereas the labels are strictly permutations. After enforcing the permutation constraint in \texttt{TAT-sequence-perm}, the performance improved across all metrics. However, the performance was still slightly lower than the baseline for most metrics. Some of the experiments in the next section were conducted in an attempt to diagnose this.

\begin{table}[htb]
    \centering
    \begin{tabular}{lcccc}
        \toprule 
        Hyperparameters & Acc/BLEU-3 & METEOR & Kendall's $\tau$ & Spearman $\rho$\\
        \midrule
        \texttt{pooling=concat, embedding\_dim=128} & 0.219 & 0.707 & 0.060 & 0.069 \\
        \texttt{pooling=mean, embedding\_dim=128} & 0.196 & 0.709 & 0.015 & 0.022 \\
        \texttt{pooling=concat, embedding\_dim=32} & 0.179 & 0.691 & 0.036 & -0.004 \\
        \bottomrule
    \end{tabular}
    \caption{Hyperparameters for sequence prediction: Results obtained by varying hyperparameters for the \texttt{TAT-sequence-perm} model using the SMS-A dataset with $n=3$.}
    \label{tab:seq-pred-hyperparam}
\end{table}

There were several experiments conducted with varied hyperparameters. The results are shown in Table \ref{tab:seq-pred-hyperparam}. The method of pooling the embeddings of the node set was varied as concatenation and mean. The embedding dimension was varied as well. The default hyperparameter choices were found to have the best performance.

\subsection{Prediction at Timestep}

\begin{table}[htb]
    \centering
    \begin{tabular}{lrccccc}
        \toprule 
        Dataset & $n$ & Acc & BLEU-3 & METEOR & Kendall's $\tau$ & Spearman $\rho$\\
        \midrule
        SMS-A & 3 & 0.263 & 0.263 & 0.690 & 0.132 & 0.141 \\
        CollegeMsg & 4 & 0.000 & 0.000 & 0.243 & 0.097 & 0.110 \\
        \bottomrule
    \end{tabular}
    \caption{Results for prediction at timestep experiments using \texttt{TAT-time-all}. CollegeMsg for $n=3$ was omitted due to compute limits.}
    \label{tab:results-pred-t-all}
\end{table}

Table \ref{tab:results-pred-t-all} shows the results for the \texttt{TAT-time-all} model. This model was developed to see whether a change in representations would improve the model's ability to learn. We see that for SMS-A for $n=3$, the model achieved higher accuracy than the baseline. Although more evidence is required, this indicates that changing the input and output space could help improve performance.

\begin{figure}[htb]
    \centering
    \includegraphics[width=0.5\linewidth]{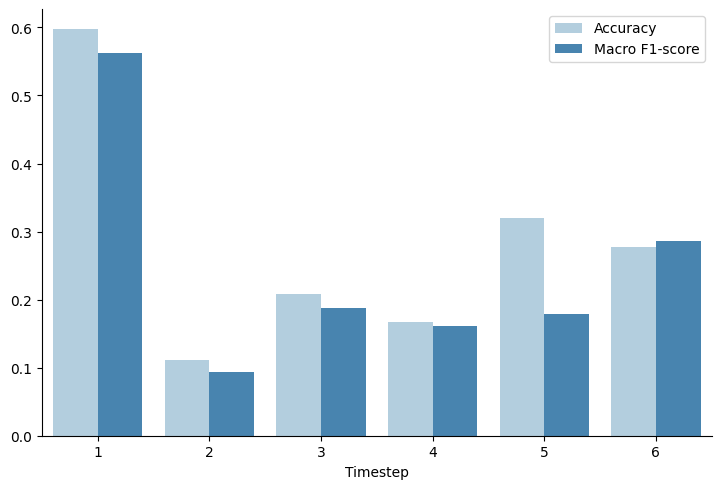}
    \caption{Results of \texttt{TAT-time-t} on CollegeMsg dataset for $n=4$ for $t=1,2,3,4,5,6$. Each point $t$ on the x-axis corresponds to the model \texttt{TAT-time-t}.}
    \label{fig:timestep_specific_predictions}
\end{figure}

Figure \ref{fig:timestep_specific_predictions} shows an example of results of \texttt{TAT-time-t} for different values of $t$. We see that the metrics are highest at time $t=1$ and drop off rapidly from the second timestep. We believe that this is because the information contained by the embeddings becomes stale over time. This motivates us to try the third set of experiments using a model which is able to learn trajectories of node embeddings so that embeddings can be projected forward in time.

\subsection{Dynamic Embeddings}

\begin{table}[htb]
    \centering
    \begin{tabular}{lcccc}
        \toprule 
        Method & Acc/BLEU-3 & METEOR & Kendall's $\tau$ & Spearman $\rho$\\
        \midrule
        \texttt{dyn-emb} & 0.175 & 0.704 & 0.023 & 0.030 \\
        \texttt{dyn-emb-projected} & 0.205 & 0.707 & 0.037 & 0.040 \\
        \bottomrule
    \end{tabular}
    \caption{Results of \texttt{dyn-emb} and \texttt{dyn-emb-projected} for the SMS-A dataset for $n=3$. The metrics averaged over 10 runs are reported.}
    \label{tab:results-jodie}
\end{table}

The results for the IOP task using the \texttt{dyn-emb} method are shown in Table \ref{tab:results-jodie}. Since the model is trained using all interactions until a certain time point, when testing on unseen interactions we expect the performance to decrease as the gap between the last timestamp seen during training and the interaction times of the test samples increases. The results for \texttt{dyn-emb} in Table \ref{tab:results-jodie} correspond to a test set consisting of $\sim$900 samples. If we used only the initial portions of the test set, the metrics were found to improve. For instance, when using the first 50, 100, 150 and 200 samples of the test set, the accuracy was found to be 0.252, 0.233, 0.205 and 0.180 respectively.

 In order to improve performance, we project the embeddings forward in time as described in Section \ref{subsec:exp-dyn-emb}. Table \ref{tab:results-jodie} shows the results for \texttt{dyn-emb-projected}. The results using projected embeddings are slightly better than using the original embeddings across all metrics.

\section{Discussion and Analysis}

The trajectory and main insights of our work can be summarized as follows. Initially, we tried two approaches to improve the baselines: the first was to use sequence prediction and the second was to use better input and output representations. Since there were no improvements to the baseline, we performed analysis and discovered that the embeddings were only useful in predicting short term interactions. This motivated us to use a better temporal embedding model. Using the new model, we observed that embeddings projected forward in time had better performance than embeddings frozen at their state at the end of training. However, due to the difference in the data used, this approach could not be compared directly with the baselines. Adjusting the frameworks to operate on the same data and evaluating them on the same data splits would be the primary direction of future work.

We also observed specific issues related to our modeling approaches, data and metrics. With the JODIE-based approach, although we had a method to project user embeddings temporally, this functionality was difficult to use. Since we do not know the timestamps at which the nodes interact, we do not know how far in time to project the embeddings. We used a naive method of projecting embeddings which provided a slight improvement in performance. There could be more sophisticated way of determining elapsed time since a user's last interaction, for example, by using the user's history of interactions. With regards to the data, we conducted experiments using only the CollegeMsg dataset when considering four-node sets. Although it had more four-node cliques than the SMS-A dataset, the number of four-node sets was still relatively low. A larger dataset would be required in order to learn better models. We evaluated experiments using the following metrics: accuracy, BLEU-3, METEOR, Kendall rank correlation coefficient and Spearman's rank correlation coefficient. In most experiments, for example in Table \ref{tab:results-seq-pred}, we observed consensus amongst the metrics. In general the metric could be chosen based on specific priorities, such as whether exact matches are important, or pairs of nodes should be placed correctly relative to each other.



\bibliographystyle{plainnat}
\bibliography{references}

\begin{thebibliography}{33}
\providecommand{\natexlab}[1]{#1}
\providecommand{\url}[1]{\texttt{#1}}
\expandafter\ifx\csname urlstyle\endcsname\relax
  \providecommand{\doi}[1]{doi: #1}\else
  \providecommand{\doi}{doi: \begingroup \urlstyle{rm}\Url}\fi

\bibitem[Chen et~al.(2019)Chen, Zhang, Xu, Fu, Zhang, Zhang, and
  Xuan]{Chen2019ELSTMDAD}
Jinyin Chen, Jian Zhang, Xuanheng Xu, Chenbo Fu, Dan Zhang, Qingpeng Zhang, and
  Qi~Xuan.
\newblock E-lstm-d: A deep learning framework for dynamic network link
  prediction.
\newblock \emph{IEEE Transactions on Systems, Man, and Cybernetics: Systems},
  51:\penalty0 3699--3712, 2019.

\bibitem[Chen et~al.(2022)Chen, Wang, and Xu]{tpl_bc2}
Jinyin Chen, Xueke Wang, and Xuanheng Xu.
\newblock {GC-LSTM}: Graph convolution embedded {LSTM} for dynamic network link
  prediction.
\newblock \emph{Applied Intelligence}, 52\penalty0 (7):\penalty0 7513--7528,
  2022.

\bibitem[Denkowski and Lavie(2014)]{meteor}
Michael Denkowski and Alon Lavie.
\newblock Meteor universal: Language specific translation evaluation for any
  target language.
\newblock In \emph{Proceedings of the EACL 2014 Workshop on Statistical Machine
  Translation}, 2014.

\bibitem[Devlin et~al.(2018)Devlin, Chang, Lee, and Toutanova]{BERT}
Jacob Devlin, Ming-Wei Chang, Kenton Lee, and Kristina Toutanova.
\newblock {BERT}: Pre-training of deep bidirectional transformers for language
  understanding, 2018.
\newblock URL \url{https://arxiv.org/abs/1810.04805}.

\bibitem[Du et~al.(2017)Du, Zhang, Wu, Moura, and Kar]{tagcn}
Jian Du, Shanghang Zhang, Guanhang Wu, Jos{\'{e}} M.~F. Moura, and Soummya Kar.
\newblock Topology adaptive graph convolutional networks.
\newblock \emph{CoRR}, abs/1710.10370, 2017.
\newblock URL \url{http://arxiv.org/abs/1710.10370}.

\bibitem[Hamilton et~al.(2017)Hamilton, Ying, and Leskovec]{graphsage}
Will Hamilton, Zhitao Ying, and Jure Leskovec.
\newblock Inductive representation learning on large graphs.
\newblock In \emph{Advances in Neural Information Processing Systems},
  volume~30. Curran Associates, Inc., 2017.

\bibitem[Hossain et~al.(2019)Hossain, Sohel, Shiratuddin, and Laga]{captioning}
MD~Zakir Hossain, Ferdous Sohel, Mohd~Fairuz Shiratuddin, and Hamid Laga.
\newblock A comprehensive survey of deep learning for image captioning.
\newblock \emph{ACM Computing Surveys (CsUR)}, 51\penalty0 (6):\penalty0 1--36,
  2019.

\bibitem[Kendall(1938)]{kendall}
M.~G. Kendall.
\newblock {A new measure of rank correlation}.
\newblock \emph{Biometrika}, 30\penalty0 (1-2):\penalty0 81--93, 06 1938.
\newblock ISSN 0006-3444.
\newblock \doi{10.1093/biomet/30.1-2.81}.
\newblock URL \url{https://doi.org/10.1093/biomet/30.1-2.81}.

\bibitem[Keskar et~al.(2019)Keskar, McCann, Varshney, Xiong, and
  Socher]{repetition-penalty}
Nitish~Shirish Keskar, Bryan McCann, Lav~R. Varshney, Caiming Xiong, and
  Richard Socher.
\newblock {CTRL:} {A} conditional transformer language model for controllable
  generation.
\newblock \emph{CoRR}, abs/1909.05858, 2019.
\newblock URL \url{http://arxiv.org/abs/1909.05858}.

\bibitem[Kipf and Welling(2017)]{gcn}
Thomas~N. Kipf and Max Welling.
\newblock Semi-supervised classification with graph convolutional networks.
\newblock In \emph{5th International Conference on Learning Representations,
  {ICLR} 2017, Toulon, France, April 24-26, 2017, Conference Track
  Proceedings}. OpenReview.net, 2017.
\newblock URL \url{https://openreview.net/forum?id=SJU4ayYgl}.

\bibitem[Kumar et~al.(2019)Kumar, Zhang, and Leskovec]{jodie}
Srijan Kumar, Xikun Zhang, and Jure Leskovec.
\newblock Predicting dynamic embedding trajectory in temporal interaction
  networks.
\newblock In \emph{Proceedings of the 25th ACM SIGKDD International Conference
  on Knowledge Discovery \& Data Mining}, KDD '19, page 1269–1278, New York,
  NY, USA, 2019. Association for Computing Machinery.
\newblock ISBN 9781450362016.
\newblock \doi{10.1145/3292500.3330895}.
\newblock URL \url{https://doi.org/10.1145/3292500.3330895}.

\bibitem[Lee and Hauskrecht(2021)]{clinicalevent}
Jeong~Min Lee and Milos Hauskrecht.
\newblock Neural clinical event sequence prediction through personalized online
  adaptive learning, 2021.
\newblock URL \url{https://arxiv.org/abs/2104.01787}.

\bibitem[Li et~al.(2022)Li, Wang, Wang, and Leskovec]{degnn}
Pan Li, Yanbang Wang, Hongwei Wang, and Jure Leskovec.
\newblock Distance encoding: Design provably more powerful neural networks for
  graph representation learning.
\newblock In \emph{Proceedings of the 34th International Conference on Neural
  Information Processing Systems}, NIPS'20, Red Hook, NY, USA, 2022. Curran
  Associates Inc.
\newblock ISBN 9781713829546.

\bibitem[Liben-Nowell and Kleinberg(2003)]{lp-for-social-network}
David Liben-Nowell and Jon Kleinberg.
\newblock The link prediction problem for social networks.
\newblock In \emph{Proceedings of the twelfth international conference on
  Information and knowledge management}, pages 556--559, 2003.

\bibitem[Nguyen et~al.(2018)Nguyen, Lee, Rossi, Ahmed, Koh, and Kim]{tlp_bc1}
Giang~Hoang Nguyen, John~Boaz Lee, Ryan~A. Rossi, Nesreen~K. Ahmed, Eunyee Koh,
  and Sungchul Kim.
\newblock Continuous-time dynamic network embeddings.
\newblock In \emph{Companion Proceedings of the The Web Conference 2018}, WWW
  '18, page 969–976, Republic and Canton of Geneva, CHE, 2018. International
  World Wide Web Conferences Steering Committee.
\newblock ISBN 9781450356404.
\newblock \doi{10.1145/3184558.3191526}.
\newblock URL \url{https://doi.org/10.1145/3184558.3191526}.

\bibitem[Panzarasa et~al.(2009)Panzarasa, Opsahl, and Carley]{collegemsg}
Pietro Panzarasa, Tore Opsahl, and Kathleen~M. Carley.
\newblock Patterns and dynamics of users' behavior and interaction: Network
  analysis of an online community.
\newblock \emph{Journal of the American Society for Information Science and
  Technology}, 60\penalty0 (5):\penalty0 911--932, 2009.
\newblock \doi{https://doi.org/10.1002/asi.21015}.
\newblock URL \url{https://onlinelibrary.wiley.com/doi/abs/10.1002/asi.21015}.

\bibitem[Papineni et~al.(2002)Papineni, Roukos, Ward, and Zhu]{bleu}
Kishore Papineni, Salim Roukos, Todd Ward, and Wei-Jing Zhu.
\newblock {BLEU}: A method for automatic evaluation of machine translation.
\newblock In \emph{Proceedings of the 40th annual meeting of the Association
  for Computational Linguistics}, pages 311--318, 2002.

\bibitem[Rossi et~al.(2021)Rossi, Barbosa, Firmani, Matinata, and
  Merialdo]{lp-for-kg}
Andrea Rossi, Denilson Barbosa, Donatella Firmani, Antonio Matinata, and Paolo
  Merialdo.
\newblock Knowledge graph embedding for link prediction: A comparative
  analysis.
\newblock \emph{ACM Transactions on Knowledge Discovery from Data (TKDD)},
  15\penalty0 (2):\penalty0 1--49, 2021.

\bibitem[Rossi et~al.(2020)Rossi, Chamberlain, Frasca, Eynard, Monti, and
  Bronstein]{tgn}
Emanuele Rossi, Ben Chamberlain, Fabrizio Frasca, Davide Eynard, Federico
  Monti, and Michael~M. Bronstein.
\newblock Temporal graph networks for deep learning on dynamic graphs.
\newblock \emph{CoRR}, abs/2006.10637, 2020.
\newblock URL \url{https://arxiv.org/abs/2006.10637}.

\bibitem[Selvarajah et~al.(2020)Selvarajah, Ragunathan, Kobti, and
  Kargar]{Selvarajah2020DynamicNL}
Kalyani Selvarajah, Kumaran Ragunathan, Ziad Kobti, and Mehdi Kargar.
\newblock Dynamic network link prediction by learning effective subgraphs using
  cnn-lstm.
\newblock \emph{2020 International Joint Conference on Neural Networks
  (IJCNN)}, pages 1--8, 2020.

\bibitem[Shi et~al.(2021)Shi, Keneshloo, Ramakrishnan, and
  Reddy]{summarization}
Tian Shi, Yaser Keneshloo, Naren Ramakrishnan, and Chandan~K. Reddy.
\newblock Neural abstractive text summarization with sequence-to-sequence
  models.
\newblock \emph{ACM/IMS Trans. Data Sci.}, 2\penalty0 (1), jan 2021.
\newblock ISSN 2691-1922.
\newblock \doi{10.1145/3419106}.
\newblock URL \url{https://doi.org/10.1145/3419106}.

\bibitem[Sun et~al.(2019)Sun, Liu, Wu, Pei, Lin, Ou, and Jiang]{bert4rec}
Fei Sun, Jun Liu, Jian Wu, Changhua Pei, Xiao Lin, Wenwu Ou, and Peng Jiang.
\newblock {BERT}4{R}ec: Sequential recommendation with bidirectional encoder
  representations from transformer, 2019.
\newblock URL \url{https://arxiv.org/abs/1904.06690}.

\bibitem[Veli{\v{c}}kovi{\'c} et~al.(2018)Veli{\v{c}}kovi{\'c}, Cucurull,
  Casanova, Romero, Li{\`o}, and Bengio]{gat}
Petar Veli{\v{c}}kovi{\'c}, Guillem Cucurull, Arantxa Casanova, Adriana Romero,
  Pietro Li{\`o}, and Yoshua Bengio.
\newblock Graph {A}ttention {N}etworks.
\newblock In \emph{International Conference on Learning Representations}, 2018.

\bibitem[Wang et~al.(2019)Wang, Zhang, Wang, Yu, Bai, Cui, and Xu]{glsm}
Yue Wang, Chenwei Zhang, Shen Wang, Philip~S. Yu, Lu~Bai, Lixin Cui, and
  Guandong Xu.
\newblock Generative temporal link prediction via self-tokenized sequence
  modeling.
\newblock \emph{CoRR}, abs/1911.11486, 2019.
\newblock URL \url{http://arxiv.org/abs/1911.11486}.

\bibitem[Wu et~al.(2020)Wu, Wu, Li, and Zhang]{Wu2020LinkPO}
Xiaomin Wu, Jianshe Wu, Yafeng Li, and Qian Zhang.
\newblock Link prediction of time-evolving network based on node ranking.
\newblock \emph{Knowl. Based Syst.}, 195:\penalty0 105740, 2020.

\bibitem[Wu et~al.(2010)Wu, Zhou, Xiao, Kurths, and Schellnhuber]{smsa}
Ye~Wu, Changsong Zhou, Jinghua Xiao, Jürgen Kurths, and Hans~Joachim
  Schellnhuber.
\newblock Evidence for a bimodal distribution in human communication.
\newblock \emph{Proceedings of the National Academy of Sciences}, 107\penalty0
  (44):\penalty0 18803--18808, 2010.
\newblock \doi{10.1073/pnas.1013140107}.
\newblock URL \url{https://www.pnas.org/doi/abs/10.1073/pnas.1013140107}.

\bibitem[Xia et~al.(2021)Xia, Li, Tian, and Li]{tat}
Wenwen Xia, Yuchen Li, Jianwei Tian, and Shenghong Li.
\newblock Forecasting interaction order on temporal graphs.
\newblock In \emph{Proceedings of the 27th ACM SIGKDD Conference on Knowledge
  Discovery \& Data Mining}, KDD '21, page 1884–1893, New York, NY, USA,
  2021. Association for Computing Machinery.
\newblock ISBN 9781450383325.
\newblock \doi{10.1145/3447548.3467341}.
\newblock URL \url{https://doi.org/10.1145/3447548.3467341}.

\bibitem[Xia et~al.(2022)Xia, Li, and Li]{tmp_timestamp}
Wenwen Xia, Yuchen Li, and Shenghong Li.
\newblock Graph neural point process for temporal interaction prediction.
\newblock \emph{IEEE Transactions on Knowledge and Data Engineering}, pages
  1--1, 2022.
\newblock \doi{10.1109/TKDE.2022.3149927}.

\bibitem[Xu et~al.(2020)Xu, Ruan, Korpeoglu, Kumar, and Achan]{tgat}
Da~Xu, Chuanwei Ruan, Evren Korpeoglu, Sushant Kumar, and Kannan Achan.
\newblock Inductive representation learning on temporal graphs.
\newblock In \emph{International Conference on Learning Representations}, 2020.
\newblock URL \url{https://openreview.net/forum?id=rJeW1yHYwH}.

\bibitem[Yang et~al.(2020)Yang, Wang, and Chu]{translation}
Shuoheng Yang, Yuxin Wang, and Xiaowen Chu.
\newblock A survey of deep learning techniques for neural machine translation.
\newblock \emph{CoRR}, abs/2002.07526, 2020.
\newblock URL \url{https://arxiv.org/abs/2002.07526}.

\bibitem[Yu et~al.(2022)Yu, Wang, and Jiang]{EDGNN}
Da~Yu, Junli Wang, and Changjun Jiang.
\newblock Neighborhood extended dynamic graph neural network.
\newblock In \emph{2022 14th International Conference on Machine Learning and
  Computing (ICMLC)}, ICMLC 2022, page 74–82, New York, NY, USA, 2022.
  Association for Computing Machinery.
\newblock ISBN 9781450395700.
\newblock \doi{10.1145/3529836.3529851}.
\newblock URL \url{https://doi.org/10.1145/3529836.3529851}.

\bibitem[Zar(2005)]{spearman}
Jerrold~H Zar.
\newblock Spearman rank correlation.
\newblock \emph{Encyclopedia of biostatistics}, 7, 2005.

\bibitem[Zheng et~al.(2022)Zheng, Wang, Wei, Liu, and Wang]{InstantGNN}
Yanping Zheng, Hanzhi Wang, Zhewei Wei, Jiajun Liu, and Sibo Wang.
\newblock Instant graph neural networks for dynamic graphs.
\newblock In \emph{Proceedings of the 28th ACM SIGKDD Conference on Knowledge
  Discovery and Data Mining}, KDD '22, page 2605–2615, New York, NY, USA,
  2022. Association for Computing Machinery.
\newblock ISBN 9781450393850.
\newblock \doi{10.1145/3534678.3539352}.
\newblock URL \url{https://doi.org/10.1145/3534678.3539352}.

\end{thebibliography}

\appendix

\section{Appendix}

\subsection{Additional Information}

\subsubsection{Equivalence of BLEU-3 and Accuracy for 3-node permutations}
\label{app:bleu-acc}
When predicting permutations, BLEU-3 is simplified to the following definition: 
$$\text{BLEU}_3 = p_1^{1/3} p_2^{1/3} p_3^{1/3}$$

where $p_1, p_2, p_3$ are 1-gram, 2-gram and 3-gram precisions. For a 3-node set BLEU-3 is equivalent to accuracy. $p_1$ is always 1 since the target and predicted sequences are permutations of each other. $p_3$ can only be 0 or 1 since there is a single trigram. If $p_3=0$ then BLEU-3 is 0, and matches accuracy which is 0 since the prediction is incorrect. If $p_3=1$ then the predicted sequence exactly matches the target sequence so $p_2$ is also 1, hence BLEU-3 and accuracy are both 1 and are equal once again.

\subsection{Other Baseline Models}

\begin{itemize}
    \item GCN: This pioneering work on GNNs introduced graph convolutional networks with layer-wise propagation using an approximation to spectral convolutions.
    \item GAT: GAT or Graph Attention Network was one of the first architectures to incorporate attention mechanisms. In this approach, self-attention weights are computed to attend to the neighbors during neighborhood aggregation.
    \item GraphSAGE: A function is learnt to sample and aggregate the local neighborhood of a node in order to generate node embeddings. A set of functions is trained rather than individual node embeddings.
    \item TAGCN: This work is a modified version of the original GCN proposed to be more theoretically sound leading to improved performance. It provides a method of designing fixed-size learnable filters whose topologies adapt to the graph. 
    \item DE-GNN: This work introduces structure related features termed distance encoding to capture spatial information regarding distances between node sets while learning representations. The proposed architecture is evaluated for link prediction and triad prediction tasks. 
    \item TAT: This model was discussed in Section \ref{sec:rel-work}.
\end{itemize}

Tables \ref{tab:baselines_n_3} and \ref{tab:baselines_n_4} show results for baseline models from the mid-report.

\label{app:baselines}
\begin{table}[htb]
    \centering
    \begin{tabular}{lllllllll}
        \toprule
        \multirow{2}{*}{Models} & \multicolumn{4}{c}{SMS-A} & \multicolumn{4}{c}{CollegeMsg} \\
        \cmidrule{2-9} 
        {} & Acc & AUC & BLEU-3 & Kendall's $\tau$ & Acc & AUC & BLEU-3 & Kendall's $\tau$ \\
        \midrule
        GCN &  0.208 &  0.578 & 0.208 & 0.115 & 0.331 &  0.752 & 0.331 & 0.347 \\
        GAT &  0.248 &  0.609 & 0.248 & 0.192 & 0.319 &  0.748 & 0.319 & 0.369 \\
        GraphSage &  0.146 & 0.560 & 0.146 & 0.179 & 0.333 &  0.767 & 0.333 & 0.319 \\
        TAGCN &  0.239 &  0.598 & 0.239 & 0.133 & 0.333 &  0.756 & 0.333 & 0.368 \\
        DE-GNN &  0.234 &  0.600 & 0.234 & 0.013 & 0.303 &  0.763 & 0.303 & 0.420 \\
        TAT &  0.257 &  0.663 & 0.257 & 0.087 & 0.392 &  0.805 & 0.392 & 0.509 \\
        \bottomrule
    \end{tabular}
    \caption{Baseline results for $n=3$.}
    \label{tab:baselines_n_3}
\end{table}

\begin{table}[H]
    \centering
    \begin{tabular}{lllll}
        \toprule
        Models & Acc & AUC & BLEU-3 & Kendall's $\tau$ \\
        \midrule
        GCN &  0.000 &  0.000 &  0.012 & 0.123  \\
        GAT &  0.000 &  0.000 &  0.024 & 0.243  \\
        GraphSage &  0.000 &  0.000 &  0.006 & 0.094  \\
        TAGCN &  0.000 &  0.000 &  0.015 & 0.313  \\
        DE-GNN &  0.000 &  0.000 &  0.009 & 0.171  \\
        TAT &  0.000 &  0.000 &  0.006 & 0.060  \\
        \bottomrule
    \end{tabular}
    \caption{Baseline results for $n=4$ for SMS-A.}
    \label{tab:baselines_n_4}
\end{table}

\end{document}